\documentclass[pdflatex,sn-basic]{sn-jnl}

\usepackage[utf8]{inputenc}
\usepackage{textcomp}

\usepackage{graphicx}
\usepackage{multirow}
\usepackage{amsmath,amssymb,amsfonts}
\usepackage{amsthm}
\usepackage{xcolor}
\usepackage{booktabs}
\usepackage{tabularx}
\usepackage{makecell}
\usepackage{caption}
\usepackage{enumitem}
\usepackage{stfloats}
\usepackage[ruled,linesnumbered]{algorithm2e}

\begin{document}

\title[Lightweight Cybersickness Detection in VR]{Lightweight Cybersickness Detection based on User-Specific Eye and Head Tracking Data in Virtual Reality}

\author*[1]{\fnm{Yijun} \sur{Wang}}\email{yijun.wang@kuleuven.be}
\author[1]{\fnm{Mihai} \sur{B\^ace}}\email{mihai.bace@kuleuven.be}
\author[1]{\fnm{Maria} \sur{Torres Vega}}\email{maria.torresvega@kuleuven.be}

\affil*[1]{\orgname{KU Leuven}, \orgaddress{\city{Leuven}, \country{Belgium}}}

\abstract{The occurrence of cybersickness in virtual reality (VR) significantly impairs users' perception and sense of immersion. Therefore, timely detection of cybersickness and the application of appropriate intervention strategies are crucial for enhancing the user experience. However, existing cybersickness detection methods often suffer from issues such as poor detection reliability across different levels of cybersickness and unnecessary model complexity. Furthermore, while cybersickness exhibits significant inter-user variability, most existing approaches aggregate all data from users and lack user-specific solutions. In this paper, we investigate a lightweight approach for cybersickness detection incorporating an ensemble learning model and user-specific eye and head tracking data. Our experiments using the open-source dataset \textit{Simulation 2021} demonstrate that feature engineering and training set construction are critical for determining detection performance. Models trained with data from similar-content segments achieve the best results, attaining detection accuracies of 93\% in the cross-user setting and 88\% in the user-personalized setting, using only 23-dimensional eye and head features. Moreover, by using user-specific data, well-tuned ensemble learning models with shorter training and inference times can be feasibly applied to real-world cybersickness detection, offering superior time efficiency and outstanding detection performance. This work offers useful evidence toward the development of lightweight and user-adaptive cybersickness detection models for VR applications.}


\maketitle

\section{Introduction}
Extended Reality (XR) is evolving at an unprecedented rate and demonstrating transformative potential across a wide range of industries, such as automotive~\cite{postelnicu2024extended}, construction~\cite{wang2018critical}, retail~\cite{plume2024exploring}, and healthcare~\cite{aruanno2025extended, garcia2020design, riva2019neuroscience}. However, users face a major problem in XR environments: Cybersickness, which impairs the immersion and continuity of the experience, and hinders the widespread adoption of XR technologies~\cite{stanney2020identifying}. Cybersickness (CS) refers to the discomfort experienced in immersive virtual environments, which is characterized by nausea, dizziness, fatigue, disorientation, and visual discomfort \cite{rebenitsch2016review}. CS creates a negative experience and even safety hazards for users, which may lead to users' negative perceptions of virtual environments and thus reduce their engagement and satisfaction. These issues motivate the development of an accurate and timely CS detection method, which could serve as an input for the early warning and mitigation of CS.

However, CS detection remains challenging due to the involvement of numerous contributing factors~\cite{sameri2024physiology}. Traditionally, this task is performed using subjective questionnaires. The Simulator Sickness Questionnaire (SSQ) \cite{kennedy1993simulator} is the most widely used questionnaire-based method, and other questionnaire-based instruments, such as the Virtual Reality Motion Sickness Questionnaire (VRSQ) \cite{kim2018virtual} and the Cybersickness in Virtual Reality Questionnaire (CSQ-VR) \cite{kourtesis2023cybersickness}, have also been developed in recent years. However, these methods do not provide real-time feedback, and lengthy questionnaires may disrupt the immersive experience and reduce users' self-perception of CS symptoms. In contrast to questionnaire-based approaches, the Fast Motion Sickness Score (FMS) ~\cite{keshavarz2011validating} provides a simple and segment-wise measure of CS, enabling in-session reporting of discomfort with minimal disruption to immersion. Nevertheless, it remains subject to reporting biases and scalability limitations. 

In this regard, objective assessment has been previously explored for autonomous CS detection~\cite{rebenitsch2016review}. Some studies have explored the use of Virtual Reality (VR) stereo video data for CS analysis and detection~\cite{islam2021cybersickness}. However, such video data primarily reflect the spatial similarity of the VR scenes, making it difficult for models to learn stable temporal features of users' internal states. Another option comes in the form of monitoring human physiology. Physiological signals such as electroencephalograms (EEG), electrocardiograms (ECG), electrodermal activity (EDA), heart rate (HR), head tracking and eye tracking have become the most commonly used signal sources in existing research \cite{rebenitsch2016review}. However, physiological signals are inherently complex and highly noisy, making them difficult to analyze. Modeling the relationship between cybersickness symptoms and physiological patterns requires methods capable of handling noisy and heterogeneous physiological signals.

To tackle this challenge, a wide range of deep learning-based methods are commonly used, including convolutional neural networks (CNNs), long short-term memory networks (LSTMs) ~\cite{islam2021cybersickness, jeong2022eyes, yalcin2024automatic}, and recurrent neural networks (RNNs) \cite{kim2019deep}, among others. However, deep learning approaches are often overly complex and over-parameterized, introducing unnecessary model complexity and increasing the risk of overfitting~\cite{kundu2023litevr}. Moreover, in model training, existing methods typically use a one-size-fits-all approach mixing data across users and content indiscriminately, ignoring the significant individual differences which CS manifests at the user level~\cite{tian2022review}. Models trained using cross-user data also tend to exhibit poor performance when applied to detecting CS of previously unseen users.

To address these issues, we present a lightweight method for in-session CS detection using only user-specific eye and head tracking data. Here, using the user-specific data refers to incorporating data from the target test user into the training process for calibration. We further categorize the target users' data into three levels according to the degree of content specificity with respect to VR segments, and experimentally investigate how different levels of specificity in training affect detection performance. To train and evaluate this method, we start from the publicly available dataset \textit{Simulation 2021}~\cite{ProcessingSim21}. We employ well-tuned ensemble learning models with optimized feature sets and user-specific calibration data, achieving detection accuracies of 93\% in the cross-user setting and 88\% in the user-personalized setting using only 23-dimensional eye and head tracking features, along with more balanced precision, recall, and F1 scores.
  
In summary, our main contributions are as follows:
\begin{itemize}

\item Our parameter-optimized ensemble learning model demonstrates that classic machine learning models can achieve higher detection performance than complex deep learning models at substantially lower computational cost, suggesting potential for practical deployment and real-time on-device inference.

\item Through ablation studies across semantic feature groups, we identify a simplified and highly relevant eye and head feature subset for CS detection, such as pupil positions, gaze origins, eye origins, and head movement parameters.

\item We define three levels of user-specific data and reveal the impact of personalized training set construction on model performance. Detection methods calibrated with user-specific data achieve better and more balanced performance.

\end{itemize}



The remainder of this paper is distributed as follows. Section~\ref{sec:Related Work} reviews related work on CS detection and personalized modeling. Section~\ref{sec:Methodology} introduces the method used in the paper, including data preprocessing, feature selection, and user-specific training data construction, and ensemble detection model. Section~\ref{sec:Experimental Setup} presents benchmark, ensemble models parametrization and hardware configuration. Section~\ref{sec:Results} reports the experimental results, including ablation studies of model, feature and user-specific training data construction, and Section~\ref{sec:Discussion} analyzes the experimental results. Finally, Section~\ref{sec:Conclusion and Future Work} concludes the paper and discusses future work.

\section{Related Work}
\label{sec:Related Work}


In recent years, physiological and behavioral signals are frequently used as objective assessments for CS detection. Sameri et al. proposed a multimodal physiological signal-based framework for cybersickness detection, which combines EEG, EDA, blood volume pulse, and skin temperature signals with machine learning algorithms, achieving an accuracy of 86.66\% in detecting elevated CS \cite{sameri2024physiology}. Qu et al. developed an LSTM-Attention neural network trained on physiological signals including EDA, ECG, and data on users' virtual avatar positions and bone rotations, enabling real-time classification and detection of CS \cite{qu2022bio}. However, most physiological signal-based approaches typically rely on additional sensing equipment, which increases the complexity of experimental implementation. Moreover, certain physiological signals such as EEG are highly susceptible to motion artifacts and environmental factors, and their complex signal patterns further complicate reliable analysis, which can ultimately degrade the stability and reliability of detection outcomes.

In contrast, eye-tracking and head-tracking signals provide rich visual perceptual information and head motion states relevant to CS, thus exhibiting significant research potential. Meanwhile, these data can be acquired directly through head-mounted displays, impose minimal interference on user interaction, and are well suited for deployment in practical VR applications. The method proposed in \cite{islam2022towards} integrated pupil diameter, gaze direction, convergence distance, head quaternion rotation and stereo image data to build a multimodal deep fusion network for modeling the evolution of CS. Jeong et al. proposed and validated a transferable attention-based multimodal deep learning framework, demonstrating that incorporating eye-tracking data can significantly enhance VR user modeling performance in cybersickness-related tasks~\cite{jeong2022eyes}. Shimada et al. employed second-level eye-related time-series indices and utilized a long short-term memory fully convolutional network to achieve four-level CS severity detection~\cite{shimada2023high}. However, existing studies remain limited in achieving computationally efficient CS detection. These approaches often rely on deep learning frameworks with highly complex architectures, which introduce unnecessary model components, prolong training time, and increase costs.

Moreover, the assessment and detection of CS is inherently a highly user-specific process~\cite{wu2025personalized}. As a result, incorporating personalization techniques often leads to improved detection performance, which has become a key focus in recent studies. The work in \cite{wu2025personalized} integrates individual attributes (e.g., age, gender, VR experience) into a Bayesian network for CS detection. Zhu et al. proposed a cross-modal joint learning framework that aligns non-invasive sensor signals (e.g., head, eye, EDA, HR) with video features to learn personalized user representations through cross-modal alignment \cite{zhu2025towards}. Tasnim et al. incorporated personal features into the model using a personalized training paradigm and designed personalized input layers and loss functions \cite{tasnim2024investigating}. However, most personalized approaches either introduce individual user attributes as variables or parameters within the model, or learn abstract representations of user characteristics through model architectural components. These approaches often do not account for users' behaviors and contextual information within specific VR content segments, whereas user- and content-level personalization is essential for a comprehensive understanding and detection of CS. Meanwhile, user-specific CS modeling methods remain underexplored and need to be further developed, together with an investigation of the signal modalities and model requirements for effective personalization.

In our study, we leverage eye-tracking and head-tracking data to extract key features and explore lightweight detection models. At the same time, we construct three levels of user-specific data with segmenting VR scene content and, through various training strategies, reveal the significant impact of user-personalized data on both detection accuracy and class balance.

\section{Methodology}
\label{sec:Methodology}

\begin{figure*}[t]
    \centering
    \includegraphics[width=0.9\linewidth]{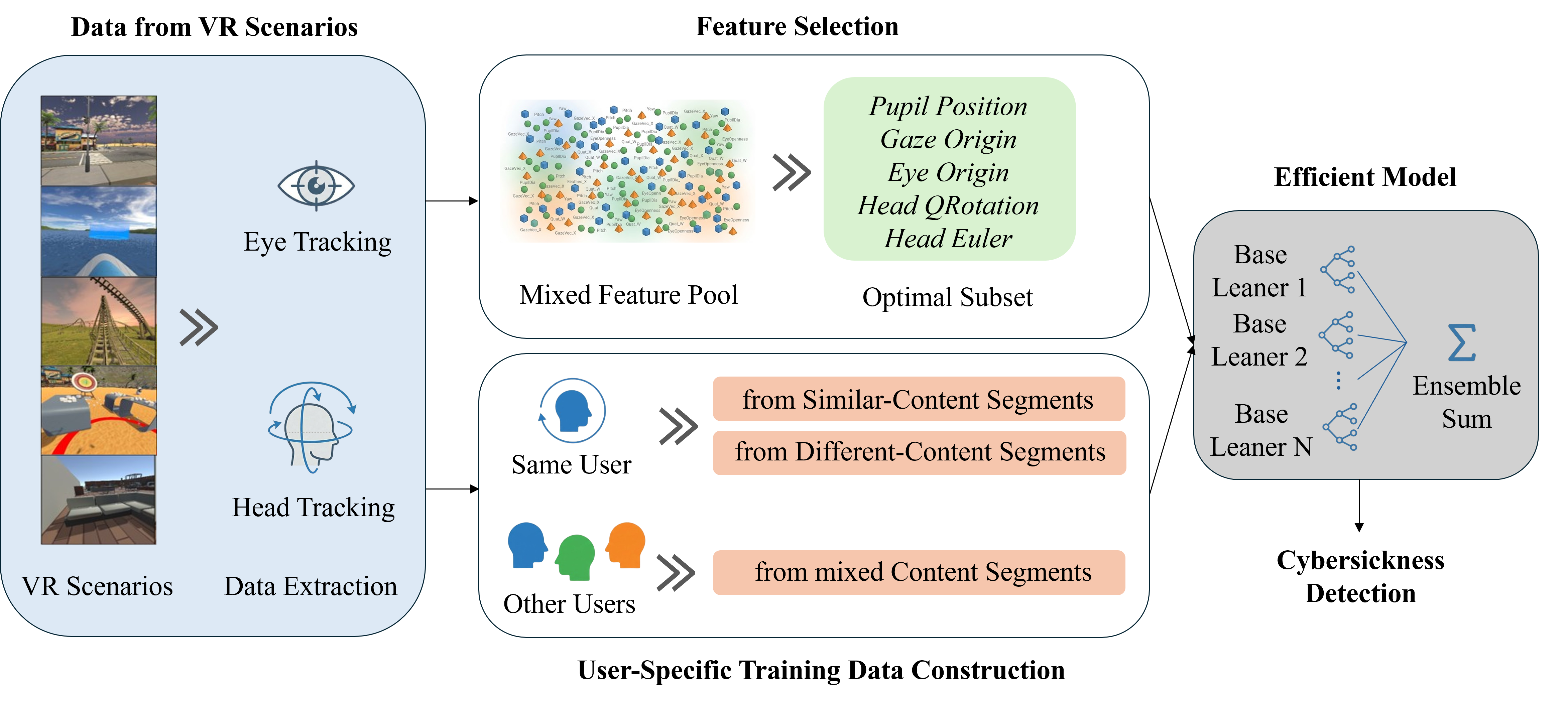}
    \caption{Overview: Eye and head tracking data extraction from VR scenarios,  feature selection, user-specific training data construction across different content segment levels, and a lightweight ensemble learning model for final CS detection.}
    \label{fig:Overview}
\end{figure*}

In this paper, we utilize eye-tracking and head-tracking signals and propose a fast and accurate CS detection framework based on feature selection, user-specific training data construction, and lightweight model design. An overview of the proposed detection methodology is illustrated in Fig.~\ref{fig:Overview}. In this section, we first introduce the dataset and the data preprocessing procedures. We then describe the extracted eye and features, the definition of the degree of data specificity, the construction of the training sets, and the ensemble detection model in turn.

\subsection{Dataset and Data Preprocessing}
\label{sec:Dataset and Data Preprocessing}
In the currently available CS datasets, resources focused on eye-tracking and head-tracking modalities remain limited. \emph{Simulation 2021} is one of the most comprehensive and frequently used datasets in this area. This dataset includes 5 VR scenarios: Road Side, SeaVoyage, Roller Coaster, Beach City, and Furniture Shop, each lasting approximately 7 minutes. In each scenario, the system played an audio prompt every 30 seconds, and participants verbally reported their current level of discomfort on the 0-10 FMS scale.

A total of 30 participants (15 males and 15 females) are included in the dataset, with a mean age of 30.04 years (standard deviation = 4.12). Among them, 19 participants had prior experience with VR devices, while 11 had no prior VR experience. No participants reported vestibular disorders or the use of medications related to motion sickness. More detailed demographic information can be found in the original papers associated with the dataset \cite{islam2021cybersickness, islam2022towards}. It is important to note that the primary objective of our study is to leverage open datasets to perform CS detection using eye and head tracking signals from head-mounted devices (HMDs), rather than to develop susceptibility models based on individual demographic factors. Therefore, demographic variables are not included as input features in our study. In the publicly released dataset, only data from 23 participants are available. Therefore, our study is based on data from these 23 participants.

To ensure comparability, our data processing pipeline follows the procedures described in the papers \cite{islam2021cybersickness, islam2022towards}. The FMS scores reported by participants are categorized into one of four CS classes: \textit{None}, \textit{Low}, \textit{Medium}, or \textit{High}. Considering the reaction delay and persistence associated with CS symptoms, a temporal buffer around the reporting time is required. We therefore construct the ground truth by aggregating an 11-second segment aligned with the FMS report, including its preceding and immediate following temporal context \cite{nalivaiko2015cybersickness, nesbitt2017correlating}. Specifically, eye and head tracking data from the segment [$D_{t-10}$, $D_{t-9}$, \ldots, $D_t$] are used to detect the CS class at time $t$. After that, outlier removal, noise reduction, and data standardization are performed.
\subsection{Feature Selection}
\label{sec:Feature Selection}
For the purpose of achieving low complexity and low sensing burden,  this work focuses on only eye tracking and head tracking data. The eye-tracking and head-tracking modalities in dataset \emph{Simulation 2021} consist of 40 dimensions in total, shown as Table~\ref{tab:used_features}.

\begin{table}[t]
    \centering
    \setlength{\tabcolsep}{3.5pt}
    \renewcommand{\arraystretch}{1.25}
    \caption{Summary of Used Eye and Head Tracking Features.}
    \label{tab:used_features}

    \begin{tabular}{@{} l @{\hspace{18pt}} l @{\hspace{18pt}} c @{}}
        \toprule
        \textbf{Variable Name} & \textbf{Additional Info} & \textbf{Dimension} \\
        \midrule
        Convergence Distance & -- & 1 \\
        \midrule
        Pupil Diameter & Both Eyes & 2 \\
        \midrule
        Eye Gaze Direction & Both Eyes; HMD Coordinate System & 6 \\
        \midrule
        Eye Openness & Both Eyes & 2 \\
        \midrule
        Pupil Position & Both Eyes; Image-Plane Coordinate System & 4 \\
        \midrule
        Eye Origin & Both Eyes; HMD Coordinate System & 6 \\
        \midrule
        Combined Gaze Origin & HMD Coordinate System & 3 \\
        \midrule
        Combined Gaze Direction & HMD Coordinate System & 3 \\
        \midrule
        Combined Gaze Origin & World Coordinate System & 3 \\
        \midrule
        Combined Gaze Direction & World Coordinate System & 3 \\
        \midrule
        Head Quaternion Rotation & -- & 4 \\
        \midrule
        Head Euler Angles & -- & 3 \\
        \bottomrule
    \end{tabular}
\end{table}

\subsection{User-Specific Training Data Construction}
\label{sec:User Specific Training Data Construction}

\begin{figure*}[t]
    \centering
    \includegraphics[width=1\linewidth]{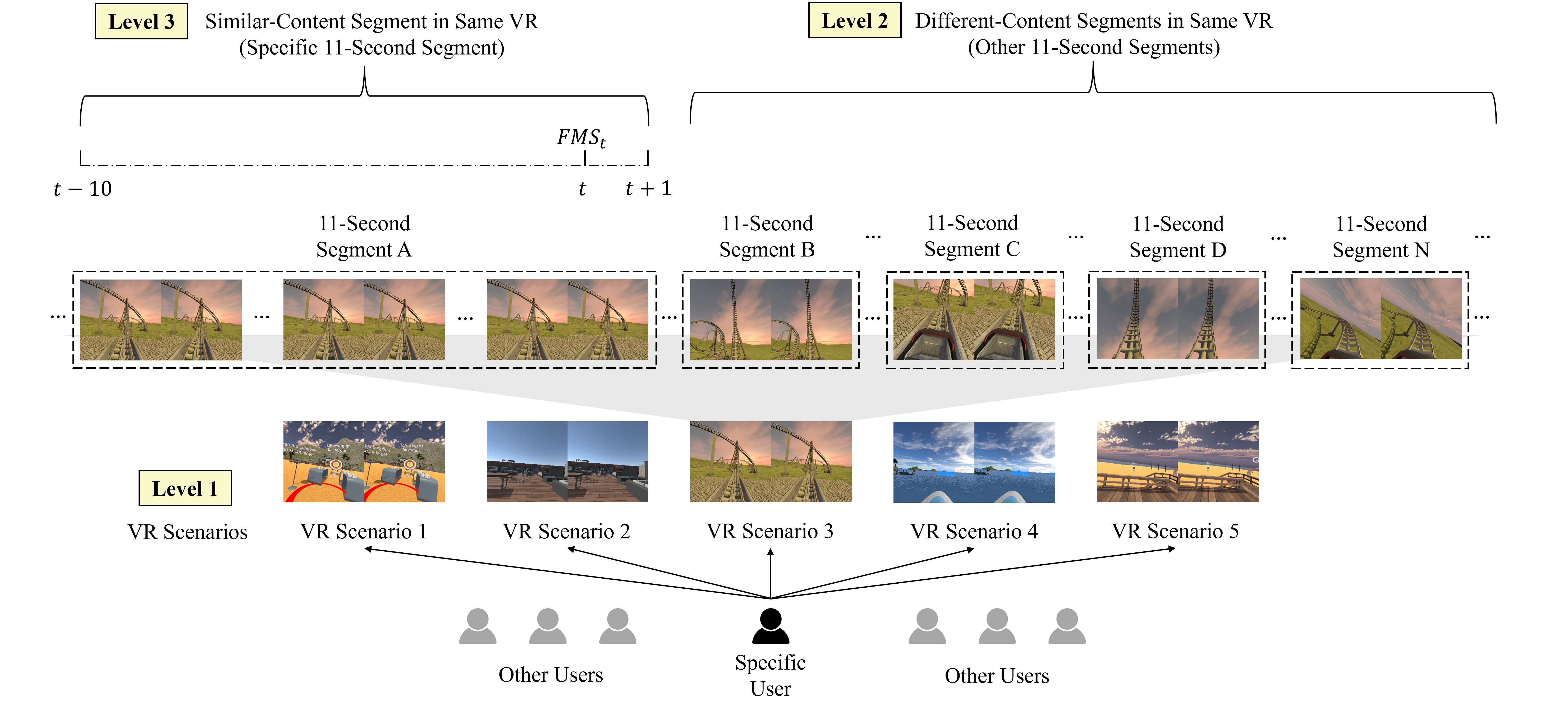}

    \caption{Three-level definition for the degree of data specificity. The 11-second segments are constructed based on the distribution of reported FMS scores. Specifically, for a given FMS score at time $t$, the data segment spanning from $t - 10$ second to $t + 1$ second is defined as a similar-content segment, while other 11-second segments are treated as different-content segments.}
    
    \label{fig:Personalized Data Definition}
\end{figure*}

Given the substantial inter-individual variability in both the manifestation and assessment of CS, user-specific data provides strong indicative value for accurate CS detection. Therefore, this work further explores the impact of user-specific input on model performance. In this context, user-specific data refers to eye and head tracking data collected from a specific user, as illustrated in Fig. \ref{fig:Personalized Data Definition}. For each user, the degree of data specificity is defined at three levels based on the associated VR content: (1) different-content segments (11-second segment) in different VR scenarios, (2) different-content segments in the same VR scenario, and (3) similar-content segment in the same VR scenario.

We should notice that learning individual biases and generalizing to new users requires a massive subject pool (typically $> 1000$) \cite{padmanaban2018towards}, which is often unavailable. So, in this paper, we adopt a cross-user mixing strategy, in which data from all users are mixed and jointly used for experiments to investigate model performance and effectiveness. This strategy is also a commonly used data-splitting strategy in CS detection studies~\cite{islam2021cybersickness, lee2019motion, padmanaban2018towards, shimada2023high}. To ensure the stability and reliability of the results, we employ 10-fold cross-validation with \emph{StratifiedKFold} function \cite{sklearn_stratifiedkfold}. During data partitioning, this function strictly maintains the class distribution within each fold to match the overall proportion of the original dataset, thereby preventing evaluation errors caused by sampling bias. Based on these settings, we further investigate the impact of the 3 levels of user-specific data on CS detection. The construction of the fixed test sets, the user-specific data, and the overall algorithmic workflow are presented in Algorithm~\ref{alg:cross_user_mixing}. Here, we extract non-overlapping 3-second windows from each 11-second segment as input samples. In each fold, the $i$-th subset $F_i$ is treated as the fixed test set, which corresponds to the samples of interest (SOI) for detection, while the remaining 9 folds are aggregated as the \textit{candidate training pool}. The \textit{candidate training pool} is then partitioned into disjoint subsets based on their relevance to the SOI: \textit{Level 1} consists of samples from unseen VR scenarios, \textit{Level 2} contains samples from seen scenarios but unseen segments, and\textit{ Level 3} includes samples from the same segments as the SOI. We construct training sets using each level individually as well as their combinations, and train and evaluate models accordingly, in order to investigate how training data with different degrees of relevance to the SOI affects detection performance.

\begin{algorithm}[t]
\caption{User-Specific Training Data Construction with 10-Fold Cross-Validation}
\label{alg:cross_user_mixing}
\SetKwInOut{Input}{Inputs}
\SetKwInOut{Output}{Outputs}
\SetKwFor{For}{for}{do}{end}

\Input{Dataset $\mathcal{D}$ (3-second samples from 11-second segments of 23 users across 5 VR scenarios)}
\Output{Mean metrics under training sets $\{L_1, L_2, L_3, L_1\!+\!L_2, L_1\!+\!L_3, L_2\!+\!L_3, L_1\!+\!L_2\!+\!L_3\}$}

\BlankLine

\tcp{Step 1: Initialization}
Partition $\mathcal{D}$ into 10 disjoint folds $\{F_1, \dots, F_{10}\}$ using \textit{StratifiedKFold}\;

\BlankLine

\tcp{Step 2: Cross-validation Loop}
\For{$i = 1$ \KwTo 10}{
    $\text{Fixed Test Set} \leftarrow F_i$\;
    $\mathcal{D}_{\text{Candidates}} \leftarrow \{ \mathcal{D} \setminus F_i \}$ (Remaining 9 folds)\;
    Identify the set of users' ID $\mathcal{U}_{test}$ present in $\text{Fixed Test Set}$\;
    
    \BlankLine
    \tcp{Step 3: User-Specific Data Construction from $\mathcal{D}_{\text{Candidates}}$}
    $\text{Level-1 User-Specific Data ($L_1$)} \leftarrow$ data in $\mathcal{D}_{\text{Candidates}}$ belonging to $\mathcal{U}_{test}$ but in \textbf{different} VR scenarios\;
    
    $\text{Level-2 User-Specific Data ($L_2$)} \leftarrow$ data in $\mathcal{D}_{\text{Candidates}}$ belonging to $\mathcal{U}_{test}$ in \textbf{same} VR scenarios but \textbf{different} 11s segments\;
    
    $\text{Level-3 User-Specific Data ($L_3$)} \leftarrow$ data in $\mathcal{D}_{\text{Candidates}}$ belonging to $\mathcal{U}_{test}$ in \textbf{same} VR scenarios and \textbf{same} 11s segments\;

    \BlankLine
   
    \tcp{Step 4: Evaluation on All Training Strategies}
    \ForEach{$T \in \{L_1, L_2, L_3, L_1\!+\!L_2, L_1\!+\!L_3, L_2\!+\!L_3, L_1\!+\!L_2\!+\!L_3\}$}{
        $\text{Metric}_T \leftarrow$ Train on $T$, test on $\text{Fixed Test Set}$ $F_i$\;
    }
    Store $\{\text{Metric}_T\}$ for the $i$-th fold\;

}

\BlankLine
\Return Overall mean of $\{\text{Metric}_{T}\}$ across all 10 folds\;
\end{algorithm}

\subsection{Ensemble Detection Model}
\label{sec:Ensemble Detection Model}
To meet the requirements of training with user-specific data, it is essential to select models that offer fast training and inference speeds while maintaining strong detection performance. Accordingly, ensemble learning is a suitable choice in this context.  By combining multiple base learners, ensemble learning enables the model  to capture data distributions from diverse perspectives,  which is particularly beneficial under small-sample settings.  This strategy effectively mitigates both bias and variance,  thereby improving robustness and generalization performance.

In this study, we employed three ensemble algorithms based on distinct strategies: Random Forest, Extra Trees, and XGBoost. The Random Forest (RF) is an ensemble method rooted in the Bagging (Bootstrap Aggregating) strategy, and the Extra Trees (ET) shares similarities with Random Forest but incorporates a higher degree of randomization during node splitting. The XGBoost (eXtreme Gradient Boosting) is a highly efficient implementation based on the Boosting strategy. Unlike Bagging, XGBoost iteratively trains weak classifiers to fit the residuals of the previous iteration.
Beyond the standalone ensembles, we further investigated a Stacking-Tree model that integrates RF, ET, and XGBoost via stacked generalization. The class-probability outputs of the three base learners are concatenated as meta-features and aggregated by a multinomial logistic regression meta-learner to yield the final prediction.

\section{Experimental Setup}
\label{sec:Experimental Setup}
This section provides first details related to the benchmarks used for the comparison of our work. Then, it discusses the parametrization of the ensemble method. Finally, it describes the hardware used for the evaluations. 
\subsection{Benchmark}
\label{sec:Benchmark}

We experimentally compare widely used deep learning models and ensemble learning in this study. Regarding deep learning approaches for CS detection, LSTM, Gated Recurrent Unit (GRU), and CNN-LSTM hybrid architectures have been adopted in the majority of prior studies and have demonstrated competitive performance. In this paper, we design the LSTM, GRU, and  CNN-LSTM structure same as \cite{kundu2025securing}. Specifically, LSTM uses 6 stacked layers with 128 hidden units; GRU follows a 3-layer design with 32/64/128 units; CNN-LSTM applies two Conv1D blocks (64 filters, kernel size 3) with two max-pooling before an LSTM layer (128 units). All models end with two fully connected layers (128 and 64 units) and a Softmax output. To mitigate overfitting, recurrent dropout is applied to recurrent layers and dropout is added before the fully connected layers. ReLU is used for all hidden fully connected layers. Beyond the LSTM, GRU, and CNN-LSTM baselines, we further benchmark three architectures following recent cybersickness studies: a time-distributed CNN-LSTM (TD-CNN-LSTM) \cite{islam2021cybersickness}, a Deep Temporal Convolutional Network (DeepTCN) \cite{tasnim2024investigating}, and an Attention-based LSTM Fully Convolutional Network (ALSTM-FCN) \cite{shimada2023high}. Specifically, TD-CNN-LSTM applies two time-distributed Conv1D layers (60/120 filters, kernel size 4) with max-pooling before an LSTM layer with 120 units, followed by a 256-unit dense layer; DeepTCN uses three residual temporal convolution blocks with dilation rates of 1, 2, and 4, followed by global average pooling and two dense layers (128 and 64 units); ALSTM-FCN combines an attention-weighted LSTM branch with 128 units and an FCN branch with three Conv1D layers (128/256/128 filters, kernel sizes 8/5/3), followed by late fusion and two dense layers (128 and 64 units). All six deep learning models are trained with Adam, batch size 256, up to 200 epochs, and early stopping (patience = 30) based on validation loss. For additional machine-learning comparison, we include an Radial Basis Function kernel support vector machine (SVM-RBF) as a baseline \cite{hadadi2022prediction}. The implemented pipeline applies standardization before classification and uses $C=1.0$, $\gamma=scale$, and probability estimation.

\subsection{Ensemble Models Parametrization}
\label{sec:Ensemble Models Parametrization}
For ensemble models, hyperparameters are selected via grid search to balance complexity and generalization. 
Random Forest uses 600 trees (max\_depth =  10, min\_samples\_split =  10, min\_samples\_leaf =  4, max\_features =  $\sqrt{N}$); 
Extra Trees uses 300 trees (max\_depth =  12, min\_samples\_split =  10, min\_samples\_leaf =  4, max\_features =  0.8). 
XGBoost uses 400 estimators (max\_depth =  6, learning rate =  0.05, subsample/colsample\_bytree =  0.8/0.8, regularization $\gamma =  0.1$ and $\alpha =  0.1$ under the multi:softprob objective). Random Forest and Extra Trees utilized deeper tree structures to capture subtle features, while XGBoost employed a lower learning rate and regularization terms to prevent overfitting.
The Stacking-Tree model uses RF, ET, and XGBoost as base learners. Their class-probability outputs are combined by a multinomial logistic-regression classifier (solver =  lbfgs, max\_iter =  2000). We use cv =  5 and stack\_method =  predict\_proba, while keeping the same hyperparameters for RF, ET, and XGBoost as in the standalone setting.

\subsection{Hardware Configuration}
\label{sec:Hardware Configuration}

All training and evaluation were performed on the Google Colab cloud computing platform. The ensemble models (RF, ET, XGBoost, and Stacking-Tree model) and SVM-RBF, computationally less intensive, were trained in a standard CPU environment (Intel Xeon @ 2.20GHz, Dual-Core) using the High-RAM runtime (approx. 25.5 GB RAM) to ensure efficient handling of high-dimensional data. The Deep learning models (GRU, CNN-LSTM, LSTM, TD-CNN-LSTM, DeepTCN, ALSTM-FCN) were deployed in a GPU-accelerated environment utilizing an NVIDIA Tesla T4 GPU (Turing architecture, 2560 CUDA cores, 320 Tensor cores, 16 GB GDDR6 VRAM).

\section{Results}
\label{sec:Results}

To ensure efficiency and optimal organization of the experimental pipeline, our experiments are conducted sequentially in the following order: (1) model comparison studies, (2) feature ablation studies, and (3) training data construction ablation studies. This design allows us to progressively identify the optimal model and the most critical feature subset, thereby enabling faster construction of personalized training data. By selecting faster and better-performing models along with compact and informative input features, we significantly reduce computational costs in subsequent experiments while ensuring the comparability and consistency of experimental results.

\subsection{Model Comparison Experiments} 
\label{sec:Model_Comparison_Experiments}
We first conducted a comparative analysis to identify the optimal model for subsequent investigations. In model comparison experiments, the complete features derived from eye and head tracking (40 dimensions) were utilized. The time window was set to 60 timestamps (corresponding to 3 seconds of data given the 20Hz sampling rate) with no overlap. To facilitate fair and clear comparison of results, we adopted the cross-user mixing with all 3 levels of user-specific data \textit{Level 1} + \textit{Level 2} + \textit{Level 3} for training data construction, and evaluated performance using 10-fold cross-validation.

To further verify the statistical significance of the performance differences among models, we adopt the Friedman test to analyze both the Accuracy and F1-Macro of all models on the 10-fold cross-validation. Specifically, within each fold all models are ranked in descending order according to the corresponding metric (1 = best), and the Friedman statistic $\chi_F^2$ is then computed to examine whether systematic differences exist among the models' average ranks:
\begin{equation}
\chi_F^2 = \frac{12N}{k(k+1)}\left[\sum_{j=1}^{k}\bar{R}_j^2 - \frac{k(k+1)^2}{4}\right],
\end{equation}
where $N = 10$ denotes the number of folds, $k$ is the number of compared models, and $\bar{R}_j$ is the average rank of model $j$ across the $N$ folds. If the resulting $p$-value is smaller than the significance level $\alpha = 0.05$, the null hypothesis that ``all models have no significant performance difference'' is rejected, indicating that statistically significant performance differences exist among the models.

Results are shown in Table \ref{tab:model_comparison}. The Friedman test yields $\chi_F^2 = 74.83$, $p = 5.14 \times 10^{-12}$ on Accuracy and $\chi_F^2 = 71.89$, $p = 1.91 \times 10^{-11}$ on F1-Macro. Since both $p$-values are far below $0.001$, the performance differences among all models on Accuracy and F1-Macro are of extremely high statistical significance and cannot be attributed to random variation across the 10 folds.

\begin{table*}[!t]
    \centering
    \caption{Model Performance Comparison. The four CS classes are None (N), Low (L), Medium (M), and High (H). Values are averaged over 10-fold cross-validation. Accuracy and F1-Macro are reported as mean $\pm$ standard deviation over the 10 folds. Avg Rank denotes the average rank across the 10 folds among all models (1 = best).}
    \label{tab:model_comparison}

    {\scriptsize
    \setlength{\tabcolsep}{3pt}
    \renewcommand{\arraystretch}{1.2}
    \centering
    \resizebox{\linewidth}{!}{%
    \begin{tabular}{@{} l c c cccc cccc cccc @{}}
        \toprule
        \textbf{Model} &
        \textbf{\makecell{Train\\Time (s)}} &
        \textbf{\makecell{Inf. Time\\/Sample (ms)}} &
        \multicolumn{4}{c}{\textbf{Precision $\uparrow$}} &
        \multicolumn{4}{c}{\textbf{Recall $\uparrow$}} &
        \multicolumn{4}{c}{\textbf{F1 $\uparrow$}} \\
        \cmidrule(lr){4-7} \cmidrule(lr){8-11} \cmidrule(lr){12-15}
        & & & N & L & M & H & N & L & M & H & N & L & M & H \\
        \midrule
        SVM-RBF        & 33.761   & 0.624 & 0.97 & 0.60 & 0.58 & 0.81 & 0.89 & 0.81 & 0.74 & 0.82 & 0.93 & 0.69 & 0.65 & 0.81 \\
        Random Forest  & 2.975   & 0.397 & 0.97 & 0.66 & 0.65 & 0.77 & 0.91 & 0.82 & 0.72 & 0.87 & 0.94 & 0.73 & 0.68 & 0.81 \\
        \textbf{XGBoost} & \textbf{3.934} & \textbf{0.049} & \textbf{0.96} & \textbf{0.78} & \textbf{0.77} & \textbf{0.86} & \textbf{0.97} & \textbf{0.78} & \textbf{0.69} & \textbf{0.87} & \textbf{0.96} & \textbf{0.78} & \textbf{0.73} & \textbf{0.86} \\
        Extra Trees    & 1.238   & 0.207 & 0.97 & 0.71 & 0.72 & 0.82 & 0.93 & 0.82 & 0.74 & 0.91 & 0.95 & 0.76 & 0.73 & 0.86 \\
        Stacking-Tree  & 279.285  & 0.630 & 0.86 & 0.93 & 0.93 & 0.92 & 0.99 & 0.43 & 0.47 & 0.63 & 0.92 & 0.59 & 0.62 & 0.75 \\
        \midrule[\heavyrulewidth]
        GRU\textsuperscript{1}        & 96.858   & 0.199 & 0.95 & 0.67 & 0.62 & 0.78 & 0.93 & 0.70 & 0.65 & 0.76 & 0.94 & 0.68 & 0.63 & 0.77 \\
        CNN-LSTM\textsuperscript{1}   & 76.805   & 0.214 & 0.94 & 0.71 & 0.69 & 0.80 & 0.95 & 0.69 & 0.66 & 0.73 & 0.94 & 0.69 & 0.67 & 0.76 \\
        LSTM\textsuperscript{1}       & 2563.467 & 0.626 & 0.95 & 0.70 & 0.64 & 0.79 & 0.94 & 0.71 & 0.65 & 0.78 & 0.94 & 0.70 & 0.64 & 0.78 \\
        TD-CNN-LSTM\textsuperscript{2}    & 119.464  & 0.201 & 0.95 & 0.71 & 0.71 & 0.81 & 0.94 & 0.74 & 0.72 & 0.75 & 0.95 & 0.72 & 0.71 & 0.77 \\
        DeepTCN\textsuperscript{3}        & 92.476   & 0.194 & 0.96 & 0.71 & 0.70 & 0.80 & 0.95 & 0.74 & 0.72 & 0.84 & 0.95 & 0.72 & 0.71 & 0.82 \\
        ALSTM-FCN\textsuperscript{4}      & 549.969  & 0.328 & 0.95 & 0.75 & 0.71 & 0.83 & 0.96 & 0.71 & 0.72 & 0.81 & 0.95 & 0.73 & 0.71 & 0.82 \\
        \bottomrule
    \end{tabular}%
    }
    \par}

    \vspace{6pt}

    {\scriptsize
    \setlength{\tabcolsep}{8pt}
    \renewcommand{\arraystretch}{1.2}
    \centering
    \scalebox{0.77}{%
    \begin{tabular}{@{} l c c c c @{}}
        \toprule
        \textbf{Model} &
        \textbf{\makecell{Acc.\\(Mean$\pm$Std) $\uparrow$}} &
        \textbf{\makecell{Acc.\\Avg Rank $\downarrow$}} &
        \textbf{\makecell{F1-Macro\\(Mean$\pm$Std) $\uparrow$}} &
        \textbf{\makecell{F1-Macro\\Avg Rank $\downarrow$}} \\
        \midrule
        SVM-RBF        & 0.86$\pm$0.02 & 10.15 & 0.77$\pm$0.03 & 7.80 \\
        Random Forest  & 0.89$\pm$0.02 & 7.15  & 0.79$\pm$0.03 & 5.60 \\
        \textbf{XGBoost} & \textbf{0.92$\pm$0.01} & \textbf{1.15} & \textbf{0.83$\pm$0.03} & \textbf{1.60} \\
        Extra Trees    & 0.90$\pm$0.01 & 3.10  & 0.82$\pm$0.02 & 2.40 \\
        Stacking-Tree  & 0.87$\pm$0.01 & 9.35  & 0.72$\pm$0.03 & 10.50 \\
        \midrule[\heavyrulewidth]
        GRU\textsuperscript{1}        & 0.87$\pm$0.02 & 8.55  & 0.75$\pm$0.03 & 9.00 \\
        CNN-LSTM\textsuperscript{1}   & 0.88$\pm$0.02 & 7.00  & 0.77$\pm$0.03 & 7.90 \\
        LSTM\textsuperscript{1}       & 0.88$\pm$0.02 & 6.90  & 0.77$\pm$0.03 & 7.30 \\
        TD-CNN-LSTM\textsuperscript{2}    & 0.89$\pm$0.02 & 5.40  & 0.79$\pm$0.02 & 6.00 \\
        DeepTCN\textsuperscript{3}        & 0.90$\pm$0.01 & 4.00  & 0.80$\pm$0.02 & 4.40 \\
        ALSTM-FCN\textsuperscript{4}      & 0.90$\pm$0.02 & 3.25  & 0.80$\pm$0.03 & 3.50 \\
        \bottomrule
    \end{tabular}%
    }
    \par}

    \begin{flushleft}
    \footnotesize\textsuperscript{1} LSTM, GRU, and CNN-LSTM follow \cite{kundu2025securing}.\\
    \footnotesize\textsuperscript{2} TD-CNN-LSTM follows \cite{islam2021cybersickness}.\\
    \footnotesize\textsuperscript{3} DeepTCN follows \cite{tasnim2024investigating}.\\
    \footnotesize\textsuperscript{4} ALSTM-FCN follows \cite{shimada2023high}.
    \end{flushleft}

\end{table*}

From the average ranks in Table \ref{tab:model_comparison}, it can be observed that XGBoost achieves average ranks of 1.15 and 1.60 on Accuracy and F1-Macro, ranking first and clearly outperforming the other models, while also maintaining relatively balanced Precision, Recall, and F1 across the CS classes. Extra Trees and ALSTM-FCN rank second and third, respectively. These results indicate that the performance advantage of ensemble learning methods (particularly XGBoost) over the baselines is both robust and statistically significant. It is worth noting that, although Stacking-Tree stacks three base ensemble learners and its training time is approximately 71 times that of XGBoost, its average ranks on Accuracy and F1-Macro are only 9.35 and 10.50, placing it in the lower-middle tier among all models. This suggests that model stacking does not yield a performance gain commensurate with its computational cost in this task.

In terms of training and inference efficiency, although the ensemble learning models run solely on CPU, their training time and per-sample inference time are overall still superior to those of the deep learning models running on GPU. Taking XGBoost as an example, compared with the best-performing deep learning model, ALSTM-FCN, XGBoost achieves a 2\% higher Accuracy and a 3\% higher F1-Macro, while requiring 139.8$\times$ less training time (3.934 seconds) and 6.7$\times$ less per-sample inference time (0.049 milliseconds). This indicates that XGBoost offers both performance and efficiency advantages in this CS detection task, making it more suitable for deployment on VR devices with limited computational resources.


\subsection{Feature Selection Experiments}
\label{sec:Feature_Selection_Experiments}

Given the superior performance of the XGBoost model, we selected it as detection model for the feature selection experiments. To systematically identify the feature subsets most relevant to CS detection, we use a semantic-group-based ablation strategy. First, the 40-dimensional feature set is divided into 10 semantic groups according to their physical meaning and signal source (see the upper part of Table~\ref{tab:feature_comparison}), where each group represents an independent eye or head signal. Next, each semantic group is used as the sole input to independently train and evaluate the model, while keeping all other experimental settings unchanged, in order to quantify its individual discriminative contribution to CS detection. Based on the accuracy results of the single-group ablation experiments, we identify the most informative feature groups. Among eye-related features, \textit{Combined Gaze Origin} (Acc. = 0.89), \textit{Eye Origin} (Acc. = 0.86), and \textit{Pupil Position} (Acc. = 0.74) significantly outperform other groups, representing the most significant indicators associated with CS. Among head-related features, \textit{Head Quaternion Rotation} (Acc. = 0.66) and \textit{Head Euler Angles} (Acc. = 0.64), as the only head motion features, provide complementary cross-modal information.

Based on these findings, we construct four feature combinations for comparison (see the lower part of Table~\ref{tab:feature_comparison}): (1) head-only features (7 dimensions), (2) eye-only high-discriminative features (16 dimensions), (3) a combination of head and eye high-discriminative features (23 dimensions), and (4) the full 40-dimensional feature set as the baseline. The results show that the 23-dimensional combination achieves superior overall performance in accuracy, precision, recall, and F1-score compared to the 40-dimensional baseline, while reducing the feature dimensionality by 42.5\%. Compared to the 16-dimensional eye-only setting, incorporating head features leads to noticeable improvements in precision and recall for minority classes (\textit{Low}, \textit{Medium}, and \textit{High}). Consequently, all subsequent experiments are conducted using this optimized 23-dimensional feature set.

\begin{table*}[t]
    \centering
    \setlength{\tabcolsep}{3.5pt}
    \renewcommand{\arraystretch}{1.25}
    \caption{Feature Subset Performance Comparison. (N: None, L: Low, M: Medium, H: High)}
    \label{tab:feature_comparison}

    \resizebox{\linewidth}{!}{%
    \begin{tabular}{@{} c l c cccc cccc cccc @{}}
        \toprule
        \textbf{\makecell{Feature\\Amount}} &
        \textbf{Features} &
        \textbf{\makecell{Acc.\\$\uparrow$}} &
        \multicolumn{4}{c}{\textbf{Precision $\uparrow$}} &
        \multicolumn{4}{c}{\textbf{Recall $\uparrow$}} &
        \multicolumn{4}{c}{\textbf{F1 $\uparrow$}} \\
        \cmidrule(lr){4-7} \cmidrule(lr){8-11} \cmidrule(lr){12-15}
        & & & N & L & M & H & N & L & M & H & N & L & M & H \\
        \midrule

        1 & Conv. Dist. &
        0.27 & 0.81 & 0.16 & 0.07 & 0.05 & 0.26 & 0.29 & 0.25 & 0.39 & 0.39 & 0.20 & 0.11 & 0.09 \\[2pt]

        2 & Eye Open. &
        0.35 & 0.78 & 0.15 & 0.07 & 0.04 & 0.38 & 0.30 & 0.20 & 0.17 & 0.51 & 0.21 & 0.10 & 0.06 \\[2pt]

        2 & Pupil Diam. &
        0.40 & 0.82 & 0.15 & 0.09 & 0.11 & 0.44 & 0.28 & 0.25 & 0.42 & 0.57 & 0.20 & 0.14 & 0.17 \\[2pt]

        4 & Pupil Pos. &
        0.74 & 0.89 & 0.40 & 0.31 & 0.45 & 0.83 & 0.46 & 0.36 & 0.65 & 0.86 & 0.43 & 0.33 & 0.53 \\[2pt]

        6 & Gaze Dir. (HMD) &
        0.63 & 0.78 & 0.21 & 0.12 & 0.12 & 0.77 & 0.24 & 0.09 & 0.13 & 0.78 & 0.22 & 0.10 & 0.13 \\[2pt]

        6 & Comb. Gaze Ori. &
        0.89 & 0.95 & 0.70 & 0.69 & 0.82 & 0.94 & 0.75 & 0.71 & 0.77 & 0.94 & 0.72 & 0.70 & 0.79 \\[2pt]

        6 & Comb. Gaze Dir. &
        0.67 & 0.81 & 0.25 & 0.20 & 0.24 & 0.80 & 0.25 & 0.17 & 0.34 & 0.81 & 0.25 & 0.18 & 0.28 \\[2pt]

        6 & Eye Ori. &
        0.86 & 0.96 & 0.63 & 0.52 & 0.72 & 0.92 & 0.71 & 0.60 & 0.77 & 0.94 & 0.66 & 0.55 & 0.74 \\[2pt]

        4 & Head QRot. &
        0.66 & 0.82 & 0.19 & 0.21 & 0.34 & 0.77 & 0.32 & 0.24 & 0.40 & 0.80 & 0.29 & 0.22 & 0.36 \\[2pt]

        3 & Head Eul. &
        0.64 & 0.81 & 0.25 & 0.16 & 0.27 & 0.76 & 0.30 & 0.19 & 0.34 & 0.78 & 0.27 & 0.17 & 0.31 \\[2pt]
        \cmidrule(l{0pt}r{0pt}){1-15}

        7 & \makecell[l]{Head QRot. + Head Eul. } &
        0.72 & 0.84 & 0.35 & 0.28 & 0.46 & 0.83 & 0.35 & 0.28 & 0.45 & 0.84 & 0.35 & 0.28 & 0.45 \\[2pt]

        16 & \makecell[l]{Pupil Pos. + Eye Ori. \\ + Comb. Gaze Ori.} &
        0.91 & 0.97 & 0.75 & 0.70 & 0.79 & 0.95 & 0.79 & 0.71 & 0.85 & 0.96 & 0.77 & 0.71 & 0.82 \\[2pt]

        \textbf{23} & \makecell[l]{\scalebox{0.9}{\textbf{Pupil Pos. + Eye Ori.}} \\ \scalebox{0.9}{\textbf{+ Comb. Gaze Ori.}} \\ \scalebox{0.9}{\textbf{+ Head QRot. + Head Eul.}}} &
        \textbf{0.92} & \textbf{0.96} & \textbf{0.78} & \textbf{0.78} & \textbf{0.85} & \textbf{0.96} & \textbf{0.81} & \textbf{0.73} & \textbf{0.85} & \textbf{0.96} & \textbf{0.79} & \textbf{0.75} & \textbf{0.85} \\[2pt]

        40 & All Eye and Head &
        0.92 & 0.96 & 0.78 & 0.77 & 0.86 & 0.97 & 0.78 & 0.69 & 0.87 & 0.96 & 0.78 & 0.73 & 0.86 \\

        \bottomrule
    \end{tabular}%
    }
\end{table*}

\subsection{User-Specific Training Experiments}
\label{sec:User-Specific Training Experiments}
As presented in Table \ref{tab:splitting_comparison}, we reports results with XGBoost for comparing different user-specific training data construction strategies.


\begin{table*}[t]
    \centering
    \setlength{\tabcolsep}{3.5pt}
    \renewcommand{\arraystretch}{1.25}
    \caption{Performance Comparison under Different Training Data Construction Strategies. (N: None, L: Low, M: Medium, H: High; Level 1: Different VR, Level 2: Different-content segments in same VR, Level 3: Similar-content segments in same VR)}
    \label{tab:splitting_comparison}

    \resizebox{\linewidth}{!}{%
    \begin{tabular}{@{} l c c cccc cccc cccc @{}}
        \toprule
        \textbf{\makecell[l]{Training Data\\Construction}} &
        \textbf{\makecell{Sample\\Size}} &
        \textbf{\makecell{Acc.\\$\uparrow$}} &
        \multicolumn{4}{c}{\textbf{Precision $\uparrow$}} &
        \multicolumn{4}{c}{\textbf{Recall $\uparrow$}} &
        \multicolumn{4}{c}{\textbf{F1 $\uparrow$}} \\
        \cmidrule(lr){4-7} \cmidrule(lr){8-11} \cmidrule(lr){12-15}
        & & & N & L & M & H & N & L & M & H & N & L & M & H \\
        \midrule

        Level 1 &
        1789 &
        0.66 & 0.76 & 0.10 & 0.10 & 0.01 & 0.86 & 0.07 & 0.05 & 0.02 & 0.81 & 0.08 & 0.07 & 0.02 \\[2pt]

        Level 2 &
        2054 &
        0.81 & 0.91 & 0.50 & 0.40 & 0.70 & 0.93 & 0.48 & 0.33 & 0.65 & 0.92 & 0.48 & 0.36 & 0.67 \\[2pt]

        \scalebox{0.9}{\textbf{Level 3}} &
        \textbf{844} &
        \textbf{0.94} & \textbf{0.96} & \textbf{0.84} & \textbf{0.88} & \textbf{0.89} & \textbf{0.98} & \textbf{0.81} & \textbf{0.79} & \textbf{0.84} & \textbf{0.97} & \textbf{0.82} & \textbf{0.83} & \textbf{0.85} \\[2pt]

        Level 1 + Level 2 &
        3843 &
        0.80 & 0.90 & 0.49 & 0.39 & 0.64 & 0.92 & 0.47 & 0.30 & 0.62 & 0.91 & 0.47 & 0.33 & 0.62 \\[2pt]

        \scalebox{0.9}{\textbf{Level 1 + Level 3}} &
        \textbf{2633} &
        \textbf{0.93} & \textbf{0.95} & \textbf{0.84} & \textbf{0.86} & \textbf{0.91} & \textbf{0.97} & \textbf{0.81} & \textbf{0.77} & \textbf{0.82} & \textbf{0.96} & \textbf{0.83} & \textbf{0.81} & \textbf{0.86} \\[2pt]

        Level 2 + Level 3 &
        2898 &
        0.92 & 0.97 & 0.79 & 0.75 & 0.87 & 0.96 & 0.78 & 0.78 & 0.88 & 0.97 & 0.79 & 0.76 & 0.86 \\[2pt]

        Level 1 + Level 2 + Level 3 &
        4687 &
        0.92 & 0.96 & 0.78 & 0.78 & 0.85 & 0.96 & 0.81 & 0.73 & 0.85 & 0.96 & 0.79 & 0.75 & 0.85 \\

        \bottomrule
    \end{tabular}%
    }
\end{table*}

\begin{figure*}[t]
    \centering
    \includegraphics[width=0.9\linewidth]{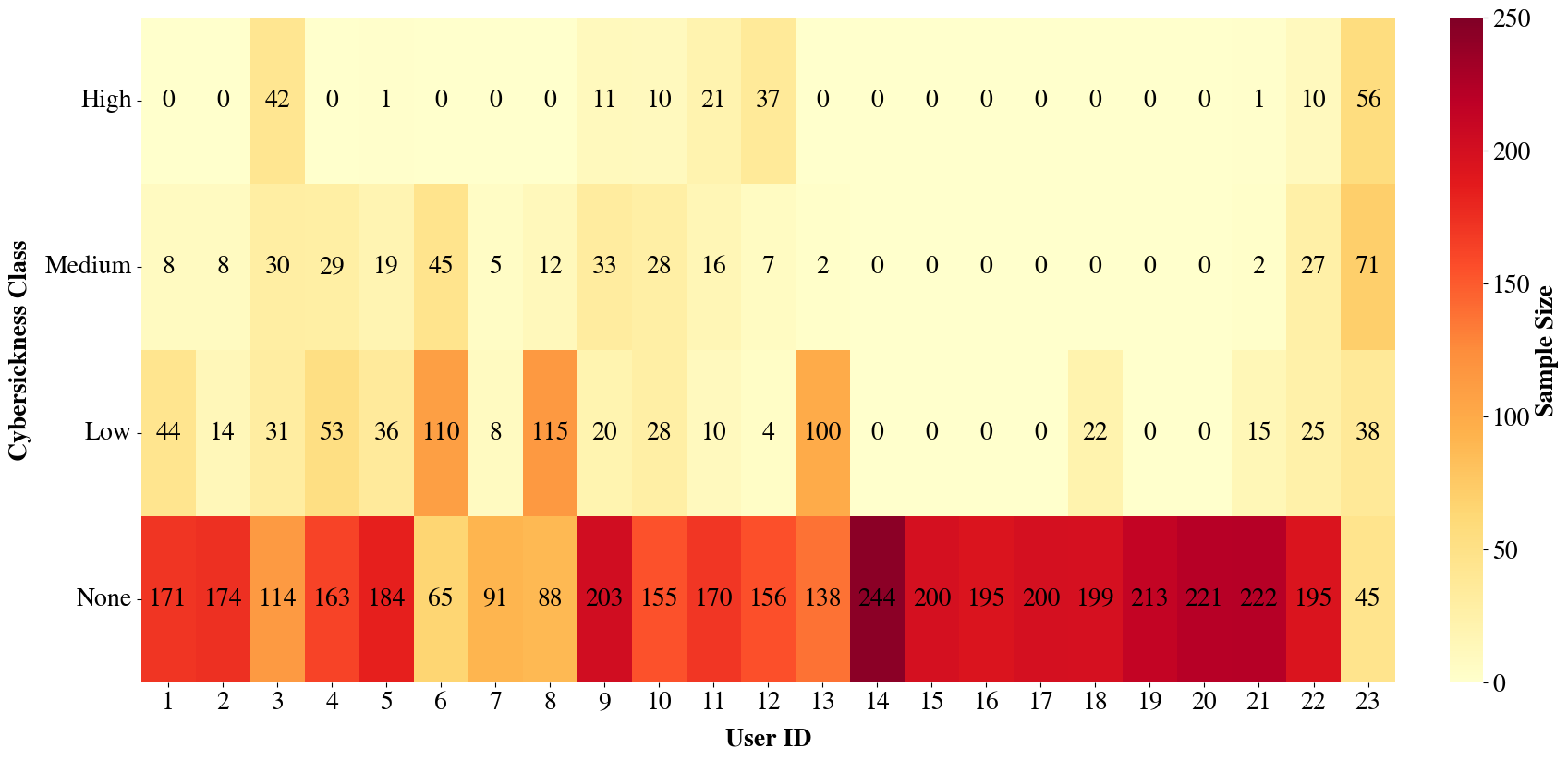}

    \caption{Per-user distribution of samples across four CS classes, illustrating the inter-user variability and severe class imbalance.}

    \label{fig:User Sample Distribution}
\end{figure*}

We can see that the training data construction with different levels of user-specific data significantly impacts experimental outcomes. Using only \textit{Level-1} user-specific data (from different content segments in different VR scenarios) has limited performance. Training solely on cross-content data limits the model's ability to learn content-specific patterns, resulting in less accurate CS detection. When using only \textit{Level-2} data (from different-content segments within the same VR scenario), performance metrics across all classes improve; however, the performance on the \textit{low} and \textit{medium} classes still exhibit relatively weak performance. Using only \textit{Level-3} data (from similar-content segments within the same VR scenario) leads to comprehensive performance gains, particularly for the minority classes such as \textit{low}, \textit{medium}, and \textit{high}. For different combinations of data levels, incorporating \textit{Level-3} training data leads to better and more balanced performance. It should be noted that using only \textit{Level-3} data corresponds to the intra-segment setting, which should be interpreted as a form of local calibration. We therefore treat its results as an optimistic upper bound of achievable performance under strong distribution alignment. In the subsequent personalized training and detection, \textit{Level-3} data are not used as independent training data, but rather as user- and content-specific local calibration. On the other hand, the \textit{Level-1 and Level-3} scheme first utilizes data from the target subject across different VR scenarios as a general base and then incorporates intra-segment data for further calibration, achieving an accuracy of 93\%. This demonstrates the feasibility of training user-specific CS detection models.

Table \ref{tab:model_comparison}, Table \ref{tab:feature_comparison}, and Table \ref{tab:splitting_comparison} report the performance of our method under the cross-user mixing strategy, consistent with \cite{islam2021cybersickness, kundu2023litevr}. This setting which mixes data from multiple users reflects overall performance on the aggregated dataset but does not reflect user-specific performance. To analyze and predict CS at the individual level, we further perform user-personalized training and CS detection for each individual user, still using XGBoost with the optimized 23-dimensional feature set. The personalized training data are constructed as \textit{Level 0 + Level 1 + Level 3}. For a selected target user, all remaining users are treated as \textit{Level 0} and serve as a general learning baseline for the model. Following the previous analysis, we adopt the most effective user-specific configuration, i.e., \textit{Level 1 + Level 3}, where \textit{Level 1} consists of the target users' data from unseen VR scenarios, and \textit{Level 3} corresponds to samples from seen VR scenarios that originate from the same segments as the SOI.

The sample distributions of different CS classes for all users in this study are illustrated in Fig.~\ref{fig:User Sample Distribution}. We observe substantial inter-user variability and severe class imbalance across users. These issues stem from extrinsic and intrinsic factors:  It is difficult to consistently induce high-level CS across all users during the experiments of the original data collection protocol~\cite{islam2021cybersickness, islam2022towards}, and individual susceptibilities to CS vary substantially. As a result, some users exhibit little or no CS symptoms throughout the sessions. For example, Users 14-17 and 19-20 contain only \textit{None} class samples, while Users 1-2, 4-8, and 21 have no or only one sample in the \textit{High} class. For these users, the scarcity of positive CS samples limits the ability of model to learn effective high-level CS patterns, and the relatively small dataset size is insufficient to support reliable zero-shot or few-shot learning.

Furthermore, these users are likely to belong to a low-susceptibility group to CS, meaning they do not exhibit high-level CS symptoms under standard VR exposure conditions. For such users, their data are predominantly concentrated in the \textit{None} or \textit{Low} classes. As a result, the model can easily learn the dominant class patterns during training and produce corresponding predictions, leading to relatively high overall performance metrics such as accuracy. However, these high metrics are largely driven by severe class imbalance. Consequently, the true impact of personalized training strategies on CS detection can not be reliably reflected, limiting the validity and interpretability of the evaluation. Therefore, here we select only representative users with relatively balanced class distributions (User 3, 9, 10, 11, 12, 22) for personalized CS training and detection. The corresponding results are reported in the Table. \ref{tab:model_per_user} and Fig.~\ref{fig:radar_per_user}.

\begin{table}[!htbp]
    \centering
    \renewcommand{\arraystretch}{1.15}
    \phantomsection
    \caption{Performance Comparison between the Baseline Training Data \textit{Level 0} and Training Data \textit{Level 0} + \textit{Level 1} + \textit{Level 3}. Results report overall accuracy and class-wise F1 scores. (N: None, L: Low, M: Medium, H: High; Level 0: Remaining users, Level 1: Different VR, Level 3: Similar-content segments in same VR)}
    \label{tab:model_per_user}

    \setlength{\tabcolsep}{11pt}
    \begin{tabular}{l l c c c c c}
        \toprule
        \textbf{\makecell[l]{Training Data\\Construction}} & \textbf{User} &
        \textbf{Acc.$\uparrow$} &
        \multicolumn{4}{c}{\textbf{F1-Score $\uparrow$}} \\
        \cmidrule(lr){4-7}
        & & & \textbf{N} & \textbf{L} & \textbf{M} & \textbf{H} \\
        \midrule

        \multirow{7}{*}{\makecell[l]{\textit{Level 0}}}
        & User 3  & 0.51 & 0.67 & 0.21 & 0.00 & 0.00 \\
        & User 9  & 0.72 & 0.84 & 0.00 & 0.00 & 0.00 \\
        & User 10 & 0.67 & 0.80 & 0.11 & 0.00 & 0.00 \\
        & User 11 & 0.74 & 0.84 & 0.00 & 0.00 & 0.00 \\
        & User 12 & 0.54 & 0.71 & 0.07 & 0.00 & 0.00 \\
        & User 22 & 0.72 & 0.84 & 0.00 & 0.18 & 0.00 \\
        \cmidrule(l){2-7}
        & \textbf{Mean} & \textbf{0.65} & \textbf{0.78} & \textbf{0.06} & \textbf{0.03} & \textbf{0.00} \\
        \midrule

        \multirow{7}{*}{\makecell[l]{\textit{Level 0} \\ + \textit{Level 1}\\+ \textit{Level 3}}}
        & User 3  & 0.83 & 0.91 & 0.68 & 0.71 & 0.67 \\
        & User 9  & 0.85 & 0.95 & 0.46 & 0.68 & 0.27 \\
        & User 10 & 0.85 & 0.92 & 0.74 & 0.65 & 0.55 \\
        & User 11 & 0.92 & 0.99 & 0.33 & 0.51 & 0.62 \\
        & User 12 & 0.95 & 0.97 & 0.40 & 0.86 & 0.94 \\
        & User 22 & 0.85 & 0.92 & 0.55 & 0.50 & 0.66 \\
        \cmidrule(l){2-7}
        & \textbf{Mean} & \textbf{0.88} & \textbf{0.94} & \textbf{0.53} & \textbf{0.65} & \textbf{0.62} \\
        \bottomrule
    \end{tabular}

    \vspace{6pt}

    {\centering\includegraphics[width=0.85\linewidth]{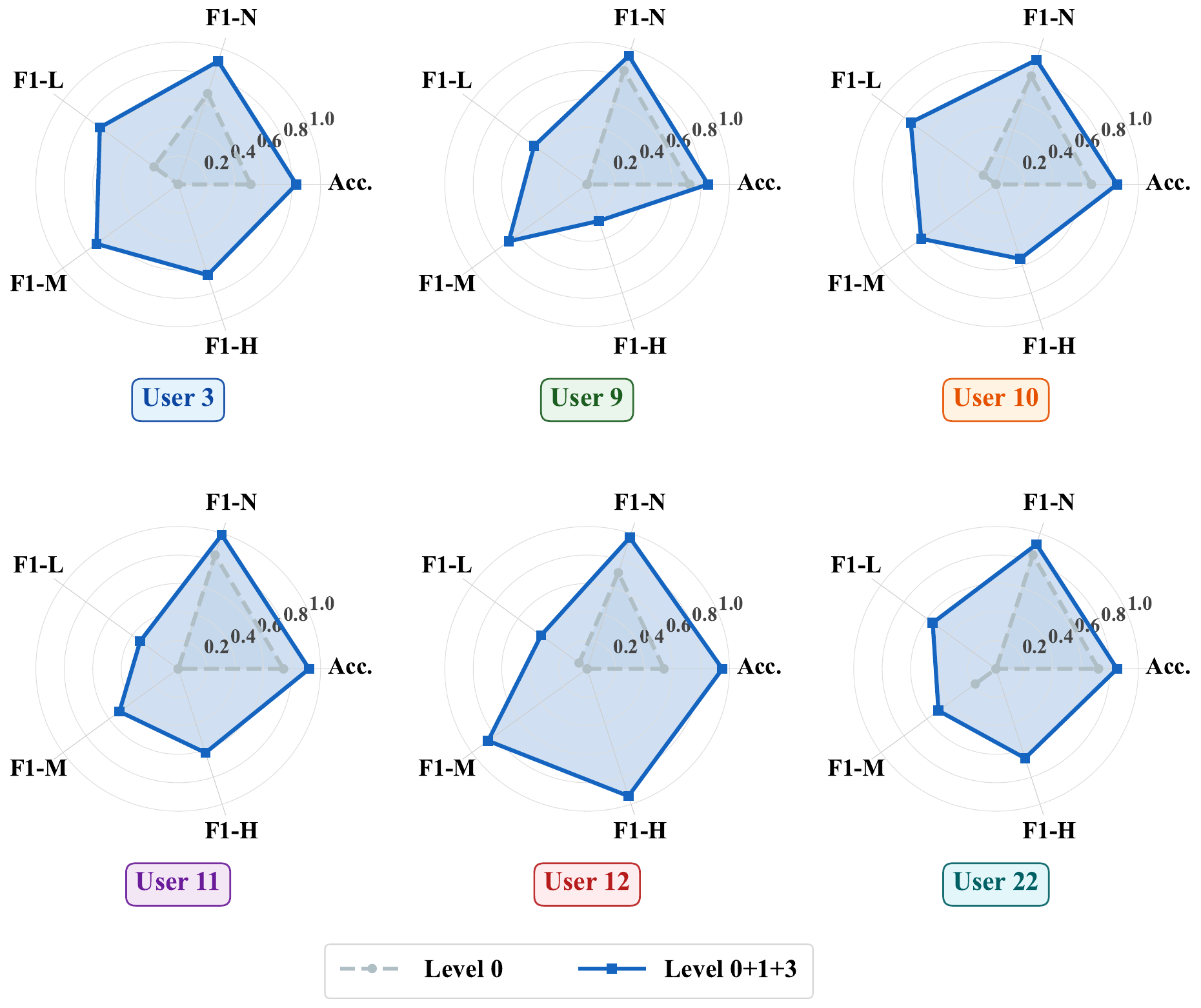}\par}
    \refstepcounter{figure}
    \par\vspace{6pt}
    {\small\textbf{Fig.~\thefigure}: Radar charts visualizing the per-user performance
    profiles from Table~\ref{tab:model_per_user}. The dashed grey
    area represents the baseline (\textit{Level~0}), and the solid
    blue area represents the personalized configuration
    (\textit{Level~0\,+\,Level~1\,+\,Level~3}).}
    \label{fig:radar_per_user}

\end{table}

As shown, when using only \textit{Level 0} for training, the average detection accuracy is merely 65\% and the model fails to effectively detect the minority classes, i.e., \textit{Low}, \textit{Medium}, and \textit{High}. After incorporating \textit{Level 1} and \textit{Level 3} and personalized training, the accuracy for each user remains high, achieving an average accuracy of 88\%, with relatively balanced performance on the minority classes. For some users, performance on specific classes (e.g., User 9 on the \textit{High} class) is lower than on other classes, which is closely related to their data distributions shown in Fig.~\ref{fig:User Sample Distribution} (e.g., User 9 has only 11 \textit{High} class samples); more balanced sample distributions generally lead to better detection results.

In terms of training efficiency, the model trained solely on \textit{Level 0} yields an average training time of 4.54 seconds and an average per-sample inference time of 0.055 milliseconds. After adding \textit{Level 1} and \textit{Level 3}, the average training time and per-sample inference time become 4.69 seconds and 0.066 milliseconds, respectively, resulting in only marginal increases of 0.15 seconds and 0.011 milliseconds. Such user-personalized detection with rapid calibration can effectively support in-session personalized detection and thus demonstrates practical deployability in real-world systems.

\section{Discussion}
\label{sec:Discussion}
In terms of CS modeling, under such small-sample and imbalanced conditions, the ensemble models under our parameter settings outperform the more complex deep learning models. In our approach, the inputs to the ensemble models are the averaged features within each window (60 timestamps). This aggregation strategy captures low-frequency variations in eye and head motions, while reducing noise and transient user-specific fluctuations. Such low-dimensional representations are well suited for robust ensemble learning models. In contrast, temporal deep learning models take the feature values at each timestep within a window as inputs, making them more prone to be affected by noise and thus more likely to suffer from overfitting. For high-dimensional inputs under label noise, inter-user variability, and limited sample sizes, the complex architectural components of deep learning models become redundant rather than beneficial. Regarding training and inference efficiency, ensemble models typically require only shallow computations. By comparison, deep learning models (e.g., CNN-LSTM) involve step-wise recurrent operations and stacked convolutions, which often run slower on GPUs than CPU-based ensemble models.

As for feature selection, based on the experimental results in Table~\ref{tab:feature_comparison}, \textit{Pupil Positions}, \textit{Gaze Origins}, \textit{Eye Origins}, \textit{Head Quaternion Rotation}, and \textit{Head Euler Angles} are identified as the most discriminative feature groups for CS detection. More specifically, \textit{Pupil Positions}, \textit{Gaze Origins}, and \textit{Eye Origins} reflect users' gaze distribution and oculomotor behavior. Gaze distribution can reflect changes in CS \cite{nam2022eye}, and oculomotor behavior (such as eye-activity features and pupil-related responses) is also linked to user-reported discomfort in VR \cite{ozkan2023relationship}. For head movement, when differences arise between the user's virtual and physical head pose due to display lag, mismatch is introduced across visual, vestibular, and proprioceptive information, which is considered one of the important triggers of CS \cite{palmisano2020cybersickness}. Accordingly, \textit{Head Quaternion Rotation} and \textit{Head Euler Angles} belong to users' head-pose dynamics, which are relevant to the magnitude and variability of this mismatch, and are therefore informative for CS detection. These eye- and head- tracking key features provide rich information about users' ocular and head activities, enabling the model to capture discriminative patterns for distinguishing different CS classes. In addition, it is important to note that a smaller set of key features can lead to better detection performance. In our multi-class setting, the boundaries between intermediate severity levels are inherently ambiguous, and the dataset is substantially imbalanced across classes. Increasing feature dimensionality without considering feature quality can make the model's precision and recall for minority classes more unstable, and may introduce additional noise or confounding information that impairs learning. 

Regarding the training data construction strategies, when additional data from similar-content segments within the same VR scenario of the specific users is introduced, this user-content representative data allows the model to build a user-tailored mapping and achieve comprehensive performance gains. Such personalized data effectively serves as an explicit form of domain adaptation and calibration, enabling the model to truly learn the users' symptom-to-signal mapping. Building on this, for single-user personalized training and detection, the \textit{Level 0 + Level 1 + Level 3} data enable the model to acquire general patterns from different users while being further calibrated with user-specific data from different VR scenarios and similar-content segments, allowing the model to capture individual user characteristics and achieve effective detection performance. This approach demonstrates favorable results in terms of overall accuracy, balanced class-wise performance, and training and inference efficiency, highlighting its practical potential for user-personalized applications.

These experimental results also provide practical insights for real-world deployment: \textit{pretrain-then-calibrate} paradigm. Data from different users can first be collected to establish a general cross-user training baseline. Subsequently, data from a specific user can be collected from different VR scenarios. These pre-collected data can be used to pretrain the model, enabling it to learn both general baseline and user prior information. As the experiment progresses, segment-specific data are continuously acquired. These user-specific data can then be progressively aggregated for calibration. Given that the lightweight ensemble model adopted in this work has low training and inference time (see Table \ref{tab:model_comparison}), the entire process can be conducted efficiently. As a result, the theoretical upper bound represented by \textit{Level-3} can be transformed into practically achievable detection performance, enabling personalized training and detection for target users in real-world applications.

\section{Conclusion and Future Work}
\label{sec:Conclusion and Future Work}

In this work, we present a lightweight in-session CS detection model with user-specific eye and head tracking data. The proposed well-tuned ensemble model, combined with content-personalized user data, achieves detection accuracies of 93\% in the cross-user setting and 88\% in the user-personalized setting using only 23-dimensional eye-tracking and head-tracking features. In this framework, during training, we combine the target users' data from different VR scenarios with content-specific data for personalized model learning. This diversified composition enables the model to learn user-specific CS patterns, leading to more accurate and balanced detection.

Our work still has several limitations. Firstly, the model training and detection in this study are partially constrained by the dataset size and sample quality. Future work should further evaluate the method's performance and robustness on larger-scale datasets with greater user diversity. Secondly, our method has not yet been deployed in a real-world system. The user-specific data required for personalized training impose relatively strict requirements on experimental design and data acquisition in practical settings. Meanwhile, in real-world applications, data loading, model training, and inference need to be performed with high computational efficiency to ensure timely and accurate CS detection.

In future work, we plan to evaluate our method under more diverse users and scenarios using finer-grained and richer eye- and head-tracking signals to further assess its real-world performance.
Meanwhile, we plan to quantify the trade-off between data size, training and calibration time, and detection performance, providing a quantitative basis for deployment under different practical conditions. In addition, the content-user personalized strategy also provides a promising path for transitioning from a general pre-trained model to a fine-tuned model using personalized data, which we will explore in the future.



\nocite{dietterich2000ensemble}
\nocite{breiman2001random}
\nocite{geurts2006extremely}
\nocite{chen2016xgboost}
\nocite{wolpert1992stacked}
\nocite{breiman1996stacked}

\bibliography{references}

\end{document}